\RequirePackage{etex}
\documentclass[conference]{IEEEtran}
\usepackage[utf8]{inputenc}

\usepackage{amsmath, amsthm, amssymb, enumitem, array}
\usepackage{algorithm, algpseudocode}
\usepackage{graphicx}
\usepackage{float}
\usepackage{listings}
\usepackage{color}
\usepackage[multiple]{footmisc}
\usepackage{multirow}
\usepackage{booktabs}
\usepackage{cite}
\usepackage{url}

\setlist[itemize]{nosep}
\setlist[enumerate]{nosep}

\definecolor{dkgreen}{rgb}{0,0.6,0}
\definecolor{gray}{rgb}{0.5,0.5,0.5}
\definecolor{mauve}{rgb}{0.58,0,0.82}

\theoremstyle{definition}

\newcolumntype{C}[1]{>{\centering\let\newline\\\arraybackslash\hspace{0pt}}m{#1}}

\begin{document}

\title{ComNetX: Local Hierarchical Adaptation for Dynamic Community Detection}

\author{
\IEEEauthorblockN{Aleksandr Konovalov, Anna Uporova, Alexander Drobyshev, Iaroslav Egorov, and Grigoriy Bokov}
\IEEEauthorblockA{AI Research Center, Lomonosov Moscow State University}
}

\maketitle

\begin{abstract}
Dynamic community detection is commonly addressed either by full-snapshot
recomputation or by solver-specific dynamic procedures. Full recomputation
preserves the semantics of mature static solvers, but it repeatedly processes
unchanged graph regions when updates are small. Solver-specific dynamic methods
can reduce this cost, but their update rules often have limited transferability
across objectives, feature representations, and implementations. In addition,
localizing computation only by graph distance may omit community context needed
by high-quality solvers. We introduce ComNetX, a solver-agnostic hierarchical
adaptation framework for local dynamic updates. ComNetX maintains a multi-level
community state, expands the updated region, closes it over affected
communities, and contracts these communities into compact local instances. This
affected-community closure and contraction preserve solver context while
restricting computation to the changed part of the graph. The same interface can
wrap modularity heuristics, graph-clustering models that use node features, and
native dynamic solvers as local backends. We evaluate ComNetX through a
multi-backend study on six real networks, longer real-data streams for
topology-based backends, and controlled dynamic stochastic block model (DSBM)
stress streams. The results show that ComNetX can preserve the quality of strong
modularity-based solvers while reducing update time on large graphs: in paired
runs on the largest real graph, Local Leiden keeps final modularity within
0.006 of full-snapshot recomputation while achieving a $41.9\pm0.2\times$
speedup. The combined protocols also identify regimes where locality breaks
down and a full refresh is preferable.
\end{abstract}



\section{Introduction}
Dynamic networks model systems whose interactions change over time, including social, communication, biological, and information networks~\cite{OSH07,PBV07,RC17}. Community detection (CD) is a central task in such data: it identifies groups of vertices with coherent structural or functional roles and supports monitoring, recommendation, anomaly detection, and exploratory analysis~\cite{POM09,SF09,RCC04,FL02}. Modularity remains one of the most widely used quality criteria~\cite{NG04,BDG08}, and scalable methods such as Louvain and Leiden are strong practical baselines~\cite{BGLL08,TWE19}; shared-memory implementations further improve static throughput~\cite{LHK14,Sah23}. However, these methods were mainly designed for static snapshots. In dynamic settings, even a small batch of edge updates may require rerunning a solver on the entire graph, which is prohibitive for large networks.

Existing dynamic CD methods reduce this cost by reusing previous partitions, applying incremental modularity updates, or processing affected neighborhoods~\cite{RC17,CT13,HKT17,ZCL19,ZK21,SLe24}. These ideas are essential for scalability, but they are often tied to a specific objective or update rule. Meanwhile, modern graph clustering increasingly incorporates attributes and neural representations~\cite{CCL23,ZXY20,DSD24,LXW26,JLL23,TPP23}. Feature-aware and GNN-based~\cite{SGH09} methods can capture richer dependencies than purely topological heuristics, yet repeatedly applying them to full snapshots is even more expensive. A dynamic framework should therefore support both classical and feature-aware solvers without redesigning each solver from scratch.
Continuous-time temporal community methods address a related but different
problem: they track communities directly in link streams and model when
vertices enter or leave them. We discuss this line in Section~\ref{S:5}, but
the target of ComNetX is batched snapshot maintenance for large dynamic graphs.

The key challenge is locality. If a batch update modifies only a small part of the graph, the algorithm should inspect and update only a bounded neighborhood of the affected vertices. This is natural for real-time graph analytics, but preserving the quality of a strong base solver under local processing is nontrivial: an overly small subgraph can lose important community context, whereas an overly large one eliminates the computational benefit.

In this paper, we present \textit{ComNetX}, a local hierarchical framework for dynamic community detection. ComNetX maintains a hierarchy of communities, identifies the communities affected by a batch update, constructs compact aggregated subproblems, and invokes the selected solver only on these local instances. The aggregation step preserves surrounding community context while avoiding full-snapshot recomputation. When a solver uses node attributes, the same aggregation pattern is applied to the feature matrix, making the framework compatible with topology-only and feature-aware methods. In this way, ComNetX acts as an algorithmic layer around existing CD solvers rather than as a solver tied to a single objective.

Our contributions are:
\begin{enumerate}[leftmargin=*]
    \item We formalize a unified dynamic adaptation setting for arbitrary community detection solvers and contrast naive full-snapshot recomputation with local processing.
    \item We introduce ComNetX, a hierarchical local wrapper that aggregates affected communities, transfers features when required, and projects updated labels back to the original graph.
    \item We analyze the update cost in terms of the affected neighborhood, the touched communities, and the contracted backend instances.
    \item We evaluate the framework with a layered protocol: a broad
    multi-backend compatibility study, longer real-data streams, and controlled
    dynamic stochastic block model (DSBM) stress streams that separate random,
    hub-centered, and
    community-internal updates.
\end{enumerate}

The paper is organized as follows. Section~\ref{S:2} gives the preliminary definitions. Section~\ref{S:3} presents the ComNetX methodology and analysis. Section~\ref{S:4} reports the experiments. Section~\ref{S:5} reviews related work, and Section~\ref{S:6} concludes the paper.

\section{Preliminaries}\label{S:2}
\subsection{Dynamic Graph Model}

Let $G_t=(V,E_t,\mathbf{A}_t,\mathbf{X})$ denote a graph snapshot at
discrete time $t$. We assume a fixed vertex universe $V$ with $n=|V|$
vertices throughout the stream. Edge updates may make a vertex isolated in a
particular snapshot, or make a previously isolated vertex incident to new
edges, but such vertices remain elements of $V$.
The weighted adjacency matrix is $\mathbf{A}_t\in\mathbb{R}^{n\times n}$.
When node attributes are available, they are stored in
$\mathbf{X}\in\mathbb{R}^{n\times d}$; otherwise the method can run with
synthetic or no features, depending on the selected backend.

The transition from $G_{t-1}$ to $G_t$ is represented by a signed batch update
$\Delta_t$:
\begin{equation*}
    \mathbf{A}_t = \mathbf{A}_{t-1} + \Delta_t .
\end{equation*}
A positive entry inserts or increases an edge weight, while a negative entry
deletes or decreases it. The endpoints of non-zero entries form the directly
affected set
\begin{equation*}
    S_t = \{\, i \in V \mid \exists j:\Delta_t(i,j)\ne 0
          \ \text{or}\ \Delta_t(j,i)\ne 0 \,\}.
\end{equation*}
For directed input graphs, ComNetX uses the symmetrized adjacency only for
detecting affected neighborhoods; the stored adjacency and the downstream
metric computation keep the representation selected by the experiment.
In the implementation, $\mathbf{A}_t$ and $\Delta_t$ are stored in sparse
matrix formats.

\subsection{Community Detection Interface}

A community detection solver is treated as a black-box procedure
\begin{equation*}
    \mathcal{B}(\mathbf{A},\mathbf{X},\mathbf{y}_0) \mapsto \mathbf{y},
\end{equation*}
where $\mathbf{A}$ is an adjacency matrix, $\mathbf{X}$ is optional, and
$\mathbf{y}_0$ is an optional warm-start partition. This interface covers
modularity heuristics such as Leiden~\cite{TWE19} and FLMIG~\cite{TTB24},
feature-aware GNN clustering models such as DMoN~\cite{TPP23},
MAGI~\cite{LLC24}, and S$^2$CAG~\cite{LYZ25}, and dynamic backends such as
MFC~\cite{KZL24} and DF-Leiden~\cite{SLe24} when they are used as local
solvers.
A \textit{naive} dynamic
adaptation of $\mathcal{B}$ simply runs it on each full snapshot:
\begin{equation*}
    \mathbf{y}_t =
    \mathcal{B}(\mathbf{A}_t,\mathbf{X},\mathbf{y}_{t-1}).
\end{equation*}
This is often a strong quality baseline, but its update time depends on the
full graph rather than on the changed region.

\subsection{Modularity}
For topology-based evaluation we use modularity~\cite{NG04}. For a weighted
adjacency matrix $\mathbf{A}$, directed or undirected, let
$W=\sum_{i,j}A_{ij}$, $d_i^{\mathrm{out}}=\sum_j A_{ij}$, and
$d_j^{\mathrm{in}}=\sum_i A_{ij}$. For partition labels $c_i$, we compute
\begin{equation*}
Q = \frac{1}{W} \sum_{i,j}
\left[A_{ij} - \gamma\frac{d_i^{\mathrm{out}}d_j^{\mathrm{in}}}{W}\right]
\mathbf{1}\{c_i=c_j\},
\end{equation*}
where $\gamma$ is the resolution parameter. For an undirected graph stored as a
symmetric adjacency, $W=2m$ and $d_i^{\mathrm{out}}=d_i^{\mathrm{in}}=d_i$,
which recovers the standard undirected definition. We use $\gamma=1$ in the
main experiments and report a resolution-sensitivity check in the ablation study.
Modularity is not a ground-truth measure and is affected by known resolution
limits~\cite{G10,FH16}; therefore we also report normalized mutual information
(NMI) as a compact external-agreement measure whenever labels are available.
For a predicted partition $C$ and ground-truth labels $Z$, we use the symmetric
normalization
\begin{equation*}
\mathrm{NMI}(C,Z)=\frac{2I(C;Z)}{H(C)+H(Z)},
\end{equation*}
where $I$ is empirical mutual information and $H$ is empirical entropy of the
corresponding label distributions~\cite{Danon05}.

\section{Methodology}\label{S:3}
\subsection{Overview}

\begin{figure*}[t]
    \centering
    \includegraphics[width=.74\textwidth,trim=0 45 0 45,clip]{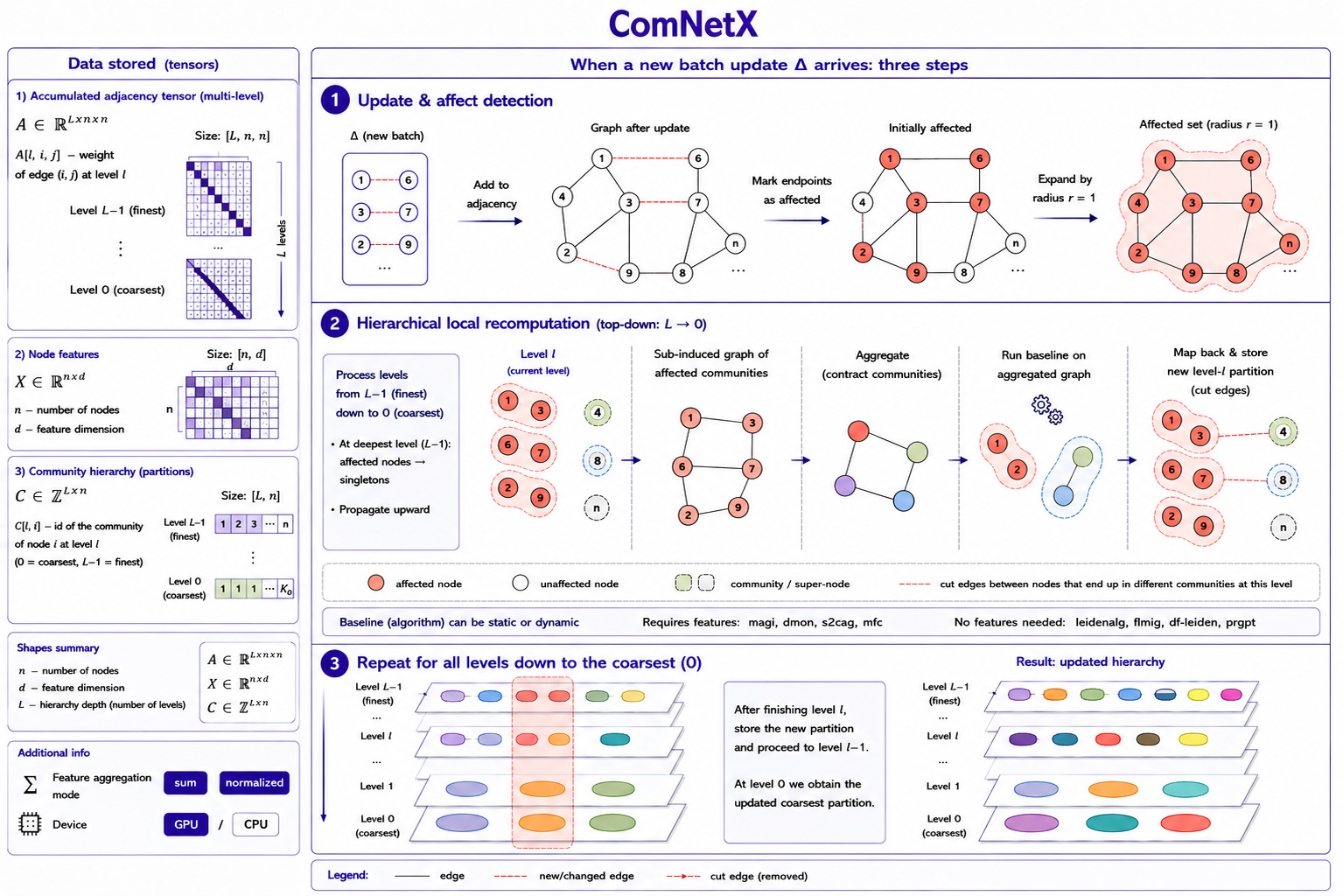}
    \caption{ComNetX update workflow: affected vertices are expanded, affected
    communities are contracted into local subproblems, the selected backend is
    run locally, and updated labels are written back to the hierarchy.}
    \label{fig:method}
\end{figure*}

ComNetX is a local adaptation layer around a base solver $\mathcal{B}$. It does
not change the objective optimized by $\mathcal{B}$. Instead, after each batch
update it constructs a compact local instance that preserves the community
context around affected vertices. The maintained state consists of
\begin{itemize}[leftmargin=*]
    \item the accumulated sparse adjacency matrix $\mathbf{A}_t$;
    \item the feature matrix $\mathbf{X}$ when the backend requires features;
    \item a hierarchy $\mathbf{C}_t\in\mathbb{Z}^{L\times n}$, where
    $\mathbf{C}_t[\ell,i]$ is the community label of vertex $i$ at hierarchy
    level $\ell$.
\end{itemize}
Figure~\ref{fig:method} summarizes the update workflow. A new batch is first
added to the accumulated graph, its endpoints are expanded to a small graph
neighborhood, and then ComNetX reoptimizes only the communities intersecting
that region. At each level, these communities are contracted into supernodes,
the base solver is executed on the contracted graph, and the result is mapped
back to the original vertices.

\subsection{Affected Region and Hierarchical Closure}

Let $r$ be a small integer radius. ComNetX expands the directly affected set
$S_t$ to
\begin{equation*}
B_r(S_t)=\{v\in V \mid \operatorname{dist}_{\mathbf{A}_t}(v,S_t)\le r\}.
\end{equation*}
Here $\operatorname{dist}_{\mathbf{A}_t}(v,S_t)=
\min_{u\in S_t}\operatorname{dist}_{\mathbf{A}_t+\mathbf{A}_t^\top}(v,u)$,
i.e., the shortest-path distance to the set $S_t$ in the graph
induced by nonzero entries of the symmetrized adjacency.

The radius-expanded affected set alone is not sufficient for high-quality local
recomputation: if only part of an existing community is included, the local
solver loses the context needed to decide whether that community should split,
merge, or remain unchanged. Therefore, at each hierarchy level $\ell$ we close
the affected set under the current community labels:
\begin{equation*}
U_t^\ell =
\{v\in V \mid \exists u\in B_r(S_t):
\mathbf{C}_{t-1}[\ell,v]=\mathbf{C}_{t-1}[\ell,u]\}.
\end{equation*}
Only vertices in $U_t^\ell$ may be relabeled at level $\ell$; vertices outside
this set keep their previous labels. Thus radius expansion identifies the
communities that may need revision, and hierarchical closure supplies the full
current context of those communities to the local backend.

\subsection{Local Aggregation}

Fix the update time $t$. At level $\ell$, let
$g_\ell(i)\in\{1,\ldots,k_\ell\}$ be the compact identifier of the current
level $\ell$ community containing $i\in U_t^\ell$. ComNetX builds a sparse
membership matrix $\mathbf{P}_\ell\in\{0,1\}^{k_\ell\times n}$:
\begin{equation*}
    \mathbf{P}_\ell[a,i]=1
    \quad\Longleftrightarrow\quad i\in U_t^\ell\ \text{and}\ g_\ell(i)=a .
\end{equation*}
Let $\mathbf{A}^{(0)}_t$ be $\mathbf{A}_t$ restricted to the affected
level 0 closure and embedded in the original $n\times n$ index space. During
the transition from level $\ell$ to level $\ell+1$, ComNetX removes the edges
of $\mathbf{A}^{(\ell)}_t$ whose endpoints belong to different level $\ell$
communities; the resulting working adjacency is $\mathbf{A}^{(\ell+1)}_t$.
The local contracted graph at level $\ell$ is then
\begin{equation*}
    \bar{\mathbf{A}}_\ell
    = \mathbf{P}_\ell \mathbf{A}^{(\ell)}_t \mathbf{P}_\ell^\top ,
\end{equation*}
so edge weights between two supernodes are exactly the sums of edge weights
between the corresponding level $\ell$ communities.

If the selected backend uses features, ComNetX uses normalized aggregation for
feature-aware local runs. Let
$n_{\ell a}=|\{i\in U_t^\ell:g_\ell(i)=a\}|$; then
\begin{equation*}
    \bar{\mathbf{X}}_\ell[a]
    =
    n_{\ell a}^{-1}\sum_{i\in U_t^\ell:g_\ell(i)=a}\mathbf{X}_i.
\end{equation*}
For topology-only solvers, feature aggregation is skipped. Structural edge
weights are preserved in $\bar{\mathbf{A}}_\ell$; in the reported feature-aware
runs they do not additionally reweight $\bar{\mathbf{X}}_\ell$.

\subsection{Relabeling and Hierarchy Update}

The contracted instance also receives a warm-start partition from the previous
hierarchy. Let $\bar{\mathbf{y}}_\ell^0[a]$ be the previous level $\ell$ label
of the vertices represented by supernode $a$, renumbered to compact labels.
Backends that do not support warm starts ignore this argument. The base solver
is invoked on the aggregated instance:
\begin{equation*}
    \bar{\mathbf{y}}_\ell =
    \mathcal{B}(\bar{\mathbf{A}}_\ell,\bar{\mathbf{X}}_\ell,\bar{\mathbf{y}}_\ell^0).
\end{equation*}
The aggregated labels are projected back to the original graph by assigning all
vertices represented by the same supernode to the corresponding new local
cluster. ComNetX reuses existing community identifiers whenever
possible, which avoids unnecessary relabeling of unchanged regions.

After updating level $\ell$, ComNetX removes from the working adjacency all
edges whose endpoints now lie in different communities at that level. This
``cut'' operation ensures that the next hierarchy level refines only the
interior of the communities produced so far. Before the level loop begins,
vertices in the expanded affected set are opened as singletons at the deepest
stored level, and affected higher level communities inherit the finer labels.
This gives the backend enough freedom to split a changed region while keeping
the unaffected part of the hierarchy fixed. Algorithm~\ref{alg:comnetx}
summarizes the update procedure for one batch.

\begin{algorithm}[t]
\caption{ComNetX update for one batch}
\label{alg:comnetx}
\begin{algorithmic}[1]
\Require accumulated graph $\mathbf{A}_{t-1}$, hierarchy $\mathbf{C}_{t-1}$,
features $\mathbf{X}$, update $\Delta_t$, radius $r$, backend $\mathcal{B}$
\State $\mathbf{A}_t \gets \mathbf{A}_{t-1}+\Delta_t$
\State $S_t \gets$ endpoints of nonzero entries in $\Delta_t$
\State $B \gets B_r(S_t)$
\State open vertices in $B$ as singletons at the deepest hierarchy level
\State propagate finer labels inside affected higher level communities
\For{each hierarchy level $\ell$}
    \State $U^\ell_t \gets$ vertices in communities intersecting $B$
    \State build sparse membership matrix $\mathbf{P}_\ell$
    \State $\bar{\mathbf{A}}_\ell \gets
        \mathbf{P}_\ell \mathbf{A}^{(\ell)}_t \mathbf{P}_\ell^\top$
    \State aggregate $\bar{\mathbf{X}}_\ell$ if $\mathcal{B}$ requires features
    \State derive warm start $\bar{\mathbf{y}}_\ell^0$ from previous level $\ell$ labels
    \State $\bar{\mathbf{y}}_\ell \gets
        \mathcal{B}(\bar{\mathbf{A}}_\ell,\bar{\mathbf{X}}_\ell,\bar{\mathbf{y}}_\ell^0)$
    \State map $\bar{\mathbf{y}}_\ell$ back to vertices in $U^\ell_t$
    \State cut inter-community edges in $\mathbf{A}^{(\ell)}_t$
\EndFor
\State \Return updated hierarchy $\mathbf{C}_t$
\end{algorithmic}
\end{algorithm}

\subsection{Complexity and Locality}

Let $b_t=|B_r(S_t)|$, let $h_\ell=|U_t^\ell|$, and let
$k_\ell$ be the number of affected communities contracted at level $\ell$.
Let $\operatorname{nnz}(\mathbf{M})$ denote the number of nonzero entries in a
sparse matrix $\mathbf{M}$. For one update, ComNetX performs neighborhood
expansion, sparse aggregation, and one backend call per level. Its time can be
written as
\begin{equation*}
\begin{aligned}
O\!\Big(&\operatorname{nnz}(\Delta_t)+\operatorname{expand}_r(S_t)\\
&+\sum_{\ell=0}^{L-1}\big[
\operatorname{nnz}(\mathbf{A}_t[U_t^\ell])
+T_{\mathcal{B}}(k_\ell,\bar m_\ell,\bar d)\big]\Big).
\end{aligned}
\end{equation*}
where $\mathbf{A}_t[U_t^\ell]$ denotes the principal submatrix of
$\mathbf{A}_t$ induced by $U_t^\ell$,
$T_{\mathcal{B}}(k_\ell,\bar m_\ell,\bar d)$ is the running time of the base
solver on the contracted instance, $\bar m_\ell=\operatorname{nnz}(\bar{\mathbf{A}}_\ell)$,
and $\bar d$ is the aggregated feature dimension when features are used. The
memory footprint of a backend call is governed by the contracted local graph,
not by the full snapshot.

For $r=1$, the expanded set satisfies
$b_t \le |S_t|+\sum_{v\in S_t} d_t(v)$, where $d_t$ is the degree in the
symmetrized expansion graph; hence hub endpoints can destroy locality even
when the number of changed edges is small. More generally, $b_t$ is controlled
by the degree profile around $S_t$ and is upper-bounded only by $|V|$ without
assumptions on degree distribution. If $\mathcal{P}_\ell$ is the previous
level $\ell$ partition, then
$h_\ell=\sum_{C\in\mathcal{P}_\ell:C\cap B_r(S_t)\ne\varnothing}|C|$ and
$k_\ell\le\min(b_t,|\mathcal{P}_\ell|)$, so closure is governed by the sizes
of the communities touched by the expanded set. Finally,
$\bar m_\ell\le\operatorname{nnz}(\mathbf{A}_t[U_t^\ell])$ because contraction
can only merge edges between the same pair of supernodes. Thus the
useful bounds are instance-dependent: in the worst case a high-degree update or
a giant affected community can make $U_t^\ell$ comparable to $V$, and ComNetX
degenerates toward full recomputation. The observable fractions $b_t/|V|$,
$\max_\ell h_\ell/|V|$, and
$\max_\ell\bar m_\ell/\operatorname{nnz}(\mathbf{A}_t)$ therefore give a
practical guardrail: if the expanded, closed, or contracted local instance
exceeds a predefined computational budget, the update should be handled by full
recomputation. The experiments below use $r=1$ and $L=3$, a configuration
selected because larger radii quickly increase the inspected region on the
studied graphs while depth values below three reduce the benefit of the
hierarchy.


\section{Experiments} \label{S:4}

This section evaluates ComNetX\footnote{\url{https://github.com/mpailab/comnetx}} as a local adapter for static and dynamic
community detection solvers. We compare it with full-snapshot recomputation
and with native dynamic baselines, focusing on the quality--time trade-off
after a sequence of graph updates.
The experiments are organized around five questions:
\begin{itemize}[leftmargin=*]
    \item \textbf{RQ1:} Does ComNetX preserve community quality while reducing
    update time relative to full-snapshot recomputation?
    \item \textbf{RQ2:} Can the same local adaptation improve or complement
    native dynamic community detection algorithms?
    \item \textbf{RQ3:} How much of the graph is inspected after neighborhood
    expansion, community closure, and contraction?
    \item \textbf{RQ4:} Which components--radius expansion, hierarchy depth,
    feature representation, closure, and contraction--control the
    quality--efficiency trade-off?
    \item \textbf{RQ5:} How stable is the method under different update streams,
    longer horizons, and adversarially large affected regions?
\end{itemize}

\subsection{Experimental setup}

\subsubsection{Datasets and Evaluation Protocols}
We evaluate on six real-world networks summarised in Table~\ref{tab:datasets}.
The selected datasets provide ground-truth labels, which allows us to complement
modularity with external clustering metrics. For this study all graphs are
converted to an undirected representation by symmetrizing the adjacency matrix.
This keeps the input format uniform across baselines, including methods whose
original implementations do not support directed updates.

The evaluation is split into the following three protocols.

\textbf{P1: multi-backend compatibility study.}
For each real network, the edge sequence is partitioned into $1000$ coarse
batches. The first $999$ batches form the initial graph $G_{999}$, and a
reference partition $P(G_{999})$ is computed with Leiden~\cite{TWE19}. The
last coarse batch is then processed as ten chronological mini-batches. This
$999{:}10$ protocol is used for broad backend coverage, including expensive
GNN baselines.

\textbf{P2: real-data batch-sensitivity and long-horizon runs.}
A strategy $p{:}q$ partitions the chronological edge stream into $p+1$ coarse
batches, builds the initial graph from the first $p$ batches, computes the
initial partition on that graph, and splits the last coarse batch into $q$
mini-batches. We use $999{:}50$ and $999{:}100$ to refine the same tail as P1
into more update steps, and $9{:}500$ to test a longer horizon from a much
earlier initial graph.

\textbf{P3: controlled DSBM stress streams.}
The synthetic protocol uses a dynamic stochastic block model (DSBM) construction
based on the stochastic block model (SBM), a standard generative model for
graphs with block-dependent edge probabilities~\cite{HLL83}. In the dynamic
setting, the graph is observed as a sequence of snapshots generated under a
time-indexed block structure~\cite{XH14}. We instantiate this idea with an
initial planted-partition graph followed by controlled update layers. The
reported 100-batch rows vary the update rate and whether changed edges are
random, hub-centered, or community-internal; different streams can therefore
end at graphs with different affected-region structure.

\begin{table}[ht]
\centering
\caption{Real-world datasets.}
\label{tab:datasets}
\begin{tabular}{lrrr}
\hline
\textbf{Dataset} & \textbf{Nodes} & \textbf{Edges} & \textbf{$Q_{\mathrm{GT}}$} \\
\hline
dyn\_cora      & 2\,708   & 5\,278   & 0.64 \\
dyn\_acm       & 3\,025   & 13\,128  & 0.48 \\
dyn\_citeseer  & 3\,327   & 4\,552   & 0.54 \\
patent         & 12\,214  & 41\,916  & 0.35 \\
dyn\_pubmed    & 19\,717  & 44\,338  & 0.43 \\
arxivmath      & 270\,013 & 799\,745 & 0.64 \\
\hline
\end{tabular}
\end{table}

The ground-truth partitions have lower modularity than the strongest
modularity-optimizing outputs on these graphs. Thus, moderate external
agreement should be interpreted together with the structural objective: high
modularity partitions may refine or reorganize label classes rather than
matching them exactly.

All wall-clock measurements were collected on a Dockerized Linux server with
two AMD EPYC 7742 64-core CPUs, 2.0 TiB RAM, and eight NVIDIA A100-SXM4 GPUs
(80 GB each). The NVIDIA driver was 535.216.03, \texttt{nvidia-smi} reported
CUDA 12.2, and the container used Python 3.10.13, PyTorch 2.3.1 with CUDA 11.8,
and TensorFlow 2.14.0.

\subsubsection{Metrics}
We evaluate both community quality and computational efficiency.
\begin{itemize}
    \item \textbf{Modularity} ($Q$): the standard modularity of the partition obtained after processing the update sequence, measured on the final graph of the corresponding protocol.
    \item \textbf{External agreement}: normalized mutual information (NMI) is computed against the ground-truth labels on the final graph of each protocol. We report NMI as a compact external metric because it is label-permutation invariant and remains comparable when the number of detected communities differs from the number of label classes.
    \item \textbf{Structural quality}: we compute final-partition mean conductance and normalized cut for full and Local Leiden on the two largest \texttt{999:10} streams. These metrics are complementary diagnostics, not optimization targets.
    \item \textbf{Total processing time} ($T$): the cumulative wall-clock time (in seconds) required to process all measured update mini-batches, i.e., the sum of the execution times of the dynamic algorithm over the whole update sequence. Data-format conversion time is excluded; each baseline operates on its native graph representation.
\end{itemize}
Whenever repeated measurements are available, comparisons are paired by dataset,
method, and batch strategy. Table~\ref{tab:main-results} gives the P1
compatibility summary, and Table~\ref{tab:stability} reports five paired
repetitions on the two largest P1 streams.

\subsubsection{Baselines}
We compare against static full-snapshot solvers, static graph-clustering
models adapted through the same snapshot interface, and native dynamic
baselines. All methods are evaluated on the same symmetrized graph sequence for
each dataset. \textit{Leidenalg}~\cite{TWE19} is the main topology-only static
baseline and represents strong modularity optimization under full
recomputation. \textit{FLMIG}~\cite{TTB24} is included as a Louvain-style local
move method with aggressive pruning. For representation-learning and
feature-aware graph clustering, we include DMoN~\cite{TPP23},
MAGI~\cite{LLC24}, S$^2$CAG~\cite{LYZ25}, and the PRGPT-Infomap and
PRGPT-Locale variants~\cite{QZG24HPEC}; DMoN, MAGI, and S$^2$CAG use node
features when available, whereas the PRGPT variants are evaluated through their
topological clustering outputs. The native dynamic baselines are
DF-Leiden~\cite{SLe24}, an OpenMP-parallel dynamic Leiden implementation, and
MFC~\cite{KZL24}, a feature-aware GNN-based dynamic community detection method.
Thus, Leidenalg, FLMIG, PRGPT-Infomap, PRGPT-Locale, and DF-Leiden are treated
as topology-only baselines in our evaluation, while DMoN, MAGI, S$^2$CAG, and
MFC are feature-aware baselines.
To keep the measurements reproducible, every empirical baseline must be
callable from the same Python-controlled batched workflow, consume the same
edge-update sequence, and return a complete hard partition for each measured
snapshot. Several related dynamic methods discussed in Sec.~\ref{S:5} are
therefore treated as Related Work rather than empirical baselines because their
available implementations require external non-Python execution paths, target
local/overlapping/continuous-time communities, or do not expose the same
full-partition output protocol. In particular, we also attempted to run the
available Java implementation of DynaMo~\cite{ZCL19} on the selected graph
streams, but it did not complete the common protocol reproducibly.

For algorithms that involve iterative training, the configured budgets are
$100$ epochs for MFC, one iteration for FLMIG, and $10$ epochs for S$^2$CAG,
MAGI, and DMoN. These values match the experimental configuration used for the
reported runs. MAGI and DMoN are not given ground-truth cluster counts. For
MAGI, after learning node embeddings, we select the number of clusters with a
vMF-mixture elbow criterion and then apply $k$-means; when a previous partition
is available, its number of labels is used only as a warm-start estimate. For
DMoN, whose output dimension is fixed in advance, we set this dimension to the
number of nodes in the current graph and compact the nonempty argmax clusters
after inference.

\textbf{Focused comparison design.}
The compatibility study covers all six real datasets and all implemented
baselines with comparable outputs. Focused tables use complete two-dataset
blocks on \textit{dyn\_pubmed} and \textit{arxivmath}. Leidenalg, DF-Leiden,
and S$^2$CAG respectively represent full recomputation, a native dynamic
topology solver, and feature-aware graph clustering under runtime pressure.
Other backends are retained in the breadth study to document compatibility and
sensitivity across a wider method family. Synthetic DSBM streams are used only
for operating-envelope stress tests, and continuous-time temporal-community
methods are treated as complementary in Related Work.

\subsubsection{Configuration of the proposed local adaptation}
Our dynamic adapter uses three hierarchy levels and a graph-neighborhood radius
of one, meaning that the endpoints of the update are expanded to their
immediate graph neighbors before community closure is applied. Separate
neighborhood-size measurements show that larger radii can quickly approach the
full graph on several datasets; this motivates the radius-one default.

Feature-aware local configurations use normalized aggregation. The reported
S$^2$CAG compatibility row uses dataset features, while
Table~\ref{tab:feature-ablation} separately evaluates dataset, random, and
one-hot representations.

\subsection{Experimental results}

\subsubsection{Compatibility Study Across Backends}

This experiment addresses RQ1.
Table~\ref{tab:main-results} summarizes runtime coverage across all protocol P1
datasets. Geometric mean speedups exclude zero-time and out-of-memory entries.
For each method, we compare the conventional baseline (``Naive'' full
recomputation or a native ``Dynamic'' solver) with the variant wrapped by the
local hierarchical adapter. In the full-snapshot setting, all arxivmath P1 rows
complete except MFC, MAGI, and DMoN; the Local adapter makes those three rows
finite as well. The long-horizon and stress tests below then assess stability
as streams lengthen or locality weakens.

\begin{table}[t]
\centering
\caption{Compatibility study across backends.}
\label{tab:main-results}
\setlength{\tabcolsep}{1.2pt}
\scriptsize
\begin{tabular}{@{}llrrrr@{}}
\toprule
\textbf{Backend} & \textbf{Base} & \textbf{$\Delta Q$} &
\textbf{$\Delta$NMI} & \textbf{\shortstack{Geometric\\mean}} &
\textbf{arxivmath} \\
\midrule
Leidenalg      & Naive   & $-0.013$ & $-0.012$ & 8.0$\times$  & 41.9$\times$ \\
FLMIG          & Naive   & $-0.247$ & $-0.039$ & 60.5$\times$ & 2606.8$\times$ \\
PRGPT-Infomap  & Naive   & $-0.131$ & $-0.064$ & 7.7$\times$  & 45.2$\times$ \\
PRGPT-Locale   & Naive   & $-0.272$ & $-0.083$ & 8.9$\times$  & 62.1$\times$ \\
S$^2$CAG       & Naive   & $+0.137$ & $+0.117$ & 5.0$\times$  & 10.2$\times$ \\
MAGI           & Naive   & $+0.081$ & $-0.035$ & 0.3$\times$  & OOM$\rightarrow$985.0s \\
DMoN           & Naive   & $+0.514$ & $+0.113$ & 1.0$\times$  & OOM$\rightarrow$90.0s \\
MFC            & Dynamic & $+0.075$ & $+0.136$ & 56.8$\times$ & OOM$\rightarrow$381.7s \\
DF-Leiden      & Dynamic & $-0.010$ & $+0.001$ & 0.1$\times$  & 8.4$\times$ \\
\bottomrule
\end{tabular}
\end{table}

\textbf{Modularity and external agreement.}
The $\Delta Q$ and $\Delta$NMI columns of Table~\ref{tab:main-results} report
aggregate P1 quality deltas, computed as the adapter result minus the baseline
result over finite baseline runs.
These values show that Local Leiden closely tracks
full-snapshot recomputation, while FLMIG and the PRGPT variants trade larger
speedups for lower aggregate quality. The feature-aware rows are more
heterogeneous: S$^2$CAG and DMoN improve both aggregate quantities under P1,
MAGI mainly contributes out-of-memory avoidance on the largest graph with mixed
NMI behavior, and MFC improves the aggregate P1 quality measures while still
remaining below the best topology-based methods in absolute quality. The
focused repeated rows in Table~\ref{tab:stability} and the S$^2$CAG
feature-mode ablation in Table~\ref{tab:feature-ablation} provide the visible
quality breakdowns for the largest datasets and feature choices.

\textbf{Reduction in processing time.}
The runtime advantage of the Local adapter is substantial for the larger
full-recomputation baselines.
For Leidenalg, Table~\ref{tab:main-results} reports a $41.9\times$ speedup on
\textit{arxivmath}; Table~\ref{tab:stability} reports the paired repeated
estimate, together with $17.1\pm0.2\times$ on \textit{dyn\_pubmed}.
FLMIG and the PRGPT variants obtain even larger or
consistent large-graph speedups, but their quality deltas show that this speed
comes from a more aggressive loss of global context. The GNN rows show a
different benefit: S$^2$CAG becomes substantially faster, while DMoN and MAGI
become executable on large graphs because only small neighborhood graphs are
materialized on the GPU; DMoN is near break-even in aggregate runtime but avoids
the full-snapshot out-of-memory failure on \textit{arxivmath}. The DF-Leiden row
shows that locality can also help an algorithm already designed for dynamic
updates when the graph is large enough.

\subsubsection{Statistical Robustness}

This experiment addresses the repeated-run part of RQ5. Table~\ref{tab:stability}
reports measurements where Local and baseline methods share the same dataset,
method settings, and batch strategy. Each row fixes the algorithm parameters;
the standard deviations therefore reflect five paired repetitions of the same
\texttt{999:10} stream shape. The Local rows use the fixed three-level,
radius-one configuration and are compared with the corresponding naive or
native dynamic baseline. Table entries use the compact mean(sd) form for
Local/Baseline quantities.

\begin{table}[t]
\centering
\caption{Repeated 999:10 robustness.}
\label{tab:stability}
\setlength{\tabcolsep}{2.5pt}
\fontsize{6.6}{7.0}\selectfont
\begin{tabular}{@{}lcc@{}}
\toprule
\textbf{Dataset} & \textbf{Q\textsubscript{L} / Q\textsubscript{B}} & \textbf{T\textsubscript{L} / T\textsubscript{B}} \\
\midrule
\multicolumn{3}{@{}l}{\textit{Leiden, Local versus full recomputation}} \\
dyn\_pubmed & 0.776\thinspace(0.000)\hspace{0.18em}/\hspace{0.18em}0.786\thinspace(0.000) & 0.39\thinspace(0.00)\hspace{0.18em}/\hspace{0.18em}6.68\thinspace(0.02) \\
arxivmath   & 0.881\thinspace(0.000)\hspace{0.18em}/\hspace{0.18em}0.886\thinspace(0.000) & 3.56\thinspace(0.02)\hspace{0.18em}/\hspace{0.18em}149.07\thinspace(0.72) \\
\addlinespace
\multicolumn{3}{@{}l}{\textit{DF-Leiden, Local versus native dynamic backend}} \\
dyn\_pubmed & 0.779\thinspace(0.000)\hspace{0.18em}/\hspace{0.18em}0.762\thinspace(0.000) & 0.24\thinspace(0.00)\hspace{0.18em}/\hspace{0.18em}0.11\thinspace(0.00) \\
arxivmath   & 0.880\thinspace(0.000)\hspace{0.18em}/\hspace{0.18em}0.850\thinspace(0.000) & 0.45\thinspace(0.01)\hspace{0.18em}/\hspace{0.18em}3.77\thinspace(0.04) \\
\addlinespace
\multicolumn{3}{@{}l}{\textit{S\textsuperscript{2}CAG, dataset features}} \\
dyn\_pubmed & 0.530\thinspace(0.000)\hspace{0.18em}/\hspace{0.18em}0.622\thinspace(0.000) & 26.44\thinspace(3.32)\hspace{0.18em}/\hspace{0.18em}111.19\thinspace(1.28) \\
arxivmath   & 0.089\thinspace(0.000)\hspace{0.18em}/\hspace{0.18em}0.001\thinspace(0.000) & 225.51\thinspace(6.82)\hspace{0.18em}/\hspace{0.18em}2297.03\thinspace(16.33) \\
\bottomrule
\end{tabular}
\end{table}

This repeated-run view shows that topology-only conclusions are stable under a
fixed stream shape: Leiden keeps a small modularity gap and large speedups, while
Local DF-Leiden is useful on \textit{arxivmath} but not on
\textit{dyn\_pubmed}, where the native dynamic backend is already efficient.
The S$^2$CAG dataset-feature rows show that the adapter can also accelerate a
feature-aware backend with real attributes: on \textit{arxivmath}, runtime drops
by $10.2\times$ and modularity improves over full-snapshot S$^2$CAG; on
\textit{dyn\_pubmed}, the speedup is smaller and the quality gap larger.

\textbf{Direction-preserving Leiden check.}
The cross-backend protocol symmetrizes real graphs because several baselines are
undirected. To check that this does not hide a Leiden-specific failure mode, we
rerun Leiden on the original directed \textit{dyn\_pubmed} and
\textit{arxivmath} streams. The directed check shows the same qualitative
pattern as the symmetrized rows: on \textit{dyn\_pubmed}, full versus Local
Leiden gives $Q=0.788$ versus $0.768$, NMI $0.208$ versus $0.199$, and
$4.99$s versus $0.36$s ($14.0\times$); on \textit{arxivmath}, the corresponding
values are $Q=0.886$ versus $0.880$, NMI $0.415$ versus $0.404$, and $108.82$s
versus $2.83$s ($38.4\times$).

\textbf{Structural-quality metrics.}
For the symmetrized \texttt{999:10} protocol used in the main real-data
comparison, we additionally compute conductance and normalized cut for full
versus Local Leiden on the two largest streams. These metrics do not cover every
backend, but they test the main topology comparison. Local Leiden reduces
normalized cut and leaves mean conductance essentially unchanged or slightly
lower:
\textit{dyn\_pubmed} changes from mean conductance/Ncut $0.128/6.17$ to
$0.127/5.33$, and \textit{arxivmath} from $0.0011/8.73$ to $0.0009/6.96$.

\subsubsection{Comparison with Native Dynamic Methods}

This experiment addresses RQ2 by comparing adapted static algorithms with native
dynamic solvers DF-Leiden and MFC on the same batched streams and final-snapshot
metrics. The comparison is an operating-envelope result, not a universal
dominance claim. In the paired \texttt{999:10} rows, Local DF-Leiden improves
modularity on both large datasets and has nearly neutral aggregate NMI, whereas
the 500-update horizon in Table~\ref{tab:long-horizon} is more conservative:
it remains faster but loses final NMI on both datasets.

DF-Leiden is already highly optimized, so the Local wrapper is not always the
fastest choice. It is slower on \textit{dyn\_pubmed}, but reaches
$8.4\pm0.1\times$ speedup on \textit{arxivmath} and remains $7.4\times$ faster
at the 500-update horizon in Table~\ref{tab:long-horizon}. MFC is slower and
runs out of memory on the largest dataset in the full dynamic setting. Thus,
native dynamic solvers remain
important speed baselines, and ComNetX is most useful when a small affected
region would otherwise trigger a much larger backend call.

\subsubsection{Locality and Workload Reduction}

This experiment addresses RQ3.
The main mechanism behind the speedup is that ComNetX transforms a full-snapshot
backend call into a sequence of small local calls. Table~\ref{tab:workload}
reports the measured growth of update neighborhoods in the real-data streams.
Since ComNetX uses $r=1$ in the main configuration, the most important column
is $|B_1(S_t)|/|V|$; the larger radii show how quickly locality would be lost
if the expansion radius were increased.
Here $|S_t|$ is the average number of directly affected vertices. The $B_h$
columns report mean vertex percentages within distance $h$ from $S_t$.

\begin{table}[!b]
\centering
\caption{Empirical neighborhood growth.}
\label{tab:workload}
\setlength{\tabcolsep}{2pt}
\scriptsize
\begin{tabular}{@{}lrrrrr@{}}
\toprule
\textbf{Dataset} & \textbf{$|S_t|$} &
\textbf{$B_1$} & \textbf{$B_2$} & \textbf{$B_3$} & \textbf{$B_4$} \\
\midrule
dyn\_cora     & 2.0   & 0.35 & 1.75 & 7.77  & 24.72 \\
dyn\_acm      & 5.2   & 2.04 & 8.11 & 21.61 & 40.60 \\
dyn\_citeseer & 2.0   & 1.01 & 3.06 & 5.82  & 10.13 \\
patent        & 5.1   & 4.49 & 5.46 & 18.57 & 18.61 \\
dyn\_pubmed   & 8.8   & 0.79 & 7.28 & 30.58 & 62.82 \\
arxivmath     & 145.6 & 0.95 & 7.01 & 27.36 & 56.60 \\
\bottomrule
\end{tabular}
\end{table}

The measured neighborhoods support $r=1$: the average radius-one region is below
$5\%$ on all datasets and below $1\%$ on \textit{dyn\_pubmed} and
\textit{arxivmath}. Larger radii quickly erode locality: $r=3$ already inspects
about $27$--$31\%$ of vertices on these two graphs, and $r=4$ exceeds half of
them. The default therefore uses immediate neighborhoods plus hierarchical
community closure rather than deeper graph-radius expansion.

\begin{figure}[!b]
\centering
\includegraphics[width=.78\columnwidth]{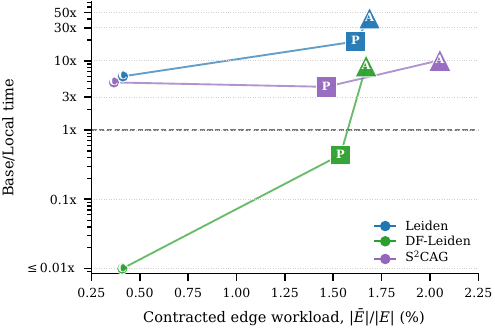}
\caption{P1 contracted workload and speedup.}
\label{fig:workload-speedup}
\end{figure}

The second locality stage is hierarchical closure and contraction, summarized
by $U_t^\ell$ and $(k_\ell,\bar m_\ell)$. For the P1 \texttt{999:10} workload
profiles, Figure~\ref{fig:workload-speedup} plots the final-level contracted
edge fraction against Base/Local time ratio, with marker area showing the mean
final-level closure fraction $|U_t^{L-1}|/|V|$ and C, P, A denoting
\textit{dyn\_cora}, \textit{dyn\_pubmed}, and \textit{arxivmath}. Backends
receive sub-percent or near-percent edge fractions, but speedups remain
backend-dependent: DF-Leiden already performs local dynamic updates, whereas
S$^2$CAG still executes neural feature materialization and tensor operations
inside each call.

The workload profiles separate locality from overhead: expansion, closure,
aggregation, and projection remain in the millisecond range, while backend
execution dominates Leiden on \textit{arxivmath} and all S$^2$CAG rows.
For S$^2$CAG, the dominant cost is the neural backend itself: feature
materialization, sparse tensor construction, and training. On A100 GPUs with
80~GB memory, full-snapshot neural runs often fit in memory, so GPU parallelism
can partly mask graph-size reduction.

\subsubsection{Ablation Study}

\begin{figure*}[!t]
\centering
\includegraphics[width=.92\textwidth]{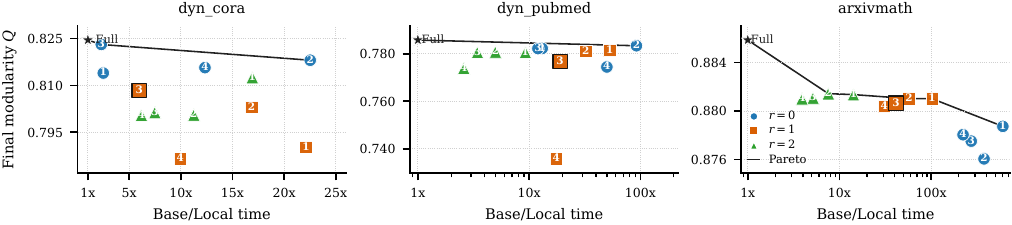}
\caption{Quality--time frontier for the Leiden topology ablation. Marker style
encodes the graph radius $r$, the marker number is the hierarchy depth $L$, and
the black line marks nondominated points.}
\label{fig:topology-ablation}
\end{figure*}

This experiment addresses RQ4.
The ablation suite isolates the tunable choices evaluated in this study:
graph-radius expansion, hierarchy depth, feature
representation, affected-community closure, and contraction. Radius
and depth are measured with Leiden to isolate the topology-only part of
ComNetX, while feature representation choices are measured with S$^2$CAG.
The closure and contraction mechanism is first quantified through
Table~\ref{tab:workload} and Figure~\ref{fig:workload-speedup}, and then tested
directly in Table~\ref{tab:closure-ablation}.
For the topology ablation, Figure~\ref{fig:topology-ablation} plots the
complete $L\times r$ grid measured for Leiden on three representative
\texttt{999:10} streams.
Each point gives final modularity $Q$ versus speedup over full-snapshot Leiden;
higher and farther right is better.

Figure~\ref{fig:topology-ablation} shows that the topology parameters are not
monotone quality knobs. On \textit{dyn\_pubmed}, endpoint-only settings dominate
the high-radius rows: $L=2,r=0$ gives the best local modularity ($Q=0.783$,
$94.5\times$), while $L=1,r=1$ remains fast ($54.4\times$) with less
quality loss than the default. On \textit{arxivmath}, the $r\ge1$ rows have
nearly the same quality ($Q\approx0.881$), but the runtime cost grows sharply
with radius; $L=1,r=0$ is much faster ($634.3\times$) with a slightly lower
$Q=0.879$. An additional \textit{dyn\_cora} grid gives yet another pattern, with
$L=3,r=0$ closest in modularity and $L=2,r=0$ the fastest near-quality point.
Together with the different neighborhood-growth rates in Table~\ref{tab:workload},
these measurements indicate that the useful $L,r$ range is topology-dependent
and that no reliable monotone selection rule was observed in this study. The
default $L=3,r=1$ is therefore a fixed cross-experiment configuration rather
than an estimated optimum. For unseen streams, a conservative budgeted protocol
is to test small-radius, shallow configurations on a pilot segment, increase
radius or depth only when quality improves while the workload fractions remain
well below the full graph, and otherwise fall back to full recomputation.

We also test whether the local contraction changes the behavior of Leiden's
resolution parameter. On the \texttt{999:50} \textit{dyn\_pubmed} and
\textit{arxivmath} streams, we rerun full and Local Leiden for
$\gamma\in\{0.5,1,2\}$. Local speedups remain large across the sweep:
$41.6$--$44.8\times$ on \textit{dyn\_pubmed} and
\mbox{$135.9$--$161.8\times$} on \textit{arxivmath}. The modularity gaps
$Q_{\mathrm{Local}}-Q_{\mathrm{Full}}$ are $-0.004$, $-0.026$, and $-0.053$
on \textit{dyn\_pubmed}, and $-0.004$, $-0.010$, and $-0.013$ on
\textit{arxivmath} for $\gamma=0.5$, $1$, and $2$, respectively. Thus the
runtime advantage is not specific to $\gamma=1$, while higher resolution can
increase quality sensitivity on some graphs. We therefore keep a fixed
resolution for paired comparisons and leave adaptive resolution selection for
local instances outside the present scope.

Table~\ref{tab:feature-ablation} reports the S$^2$CAG feature-mode ablation
under the P2 \texttt{999:100} strategy. Each cell is final modularity $Q$,
cumulative time $T$, and final-snapshot NMI. The compact random feature Local
row is close to the dataset-feature Local row while being cheaper, which makes
it a reasonable default when the original attributes are unavailable or costly.

\begin{table}[H]
\centering
\caption{S$^2$CAG 999:100 feature modes.}
\label{tab:feature-ablation}
\setlength{\tabcolsep}{2pt}
\scriptsize
\begin{tabular}{@{}lcc@{}}
\toprule
\textbf{Variant} & \textbf{dyn\_cora} & \textbf{dyn\_pubmed} \\
 & \textbf{Q / T / NMI} & \textbf{Q / T / NMI} \\
\midrule
Naive, dataset features & 0.691 / 35.71 / 0.437 & 0.624 / 372.45 / 0.202 \\
Naive, random features  & $-$0.001 / 13.02 / 0.023 & $-$0.002 / 92.18 / 0.000 \\
Naive, one-hot features & 0.751 / 43.96 / 0.360 & 0.528 / 8568.07 / 0.059 \\
Local, dataset features & 0.788 / 6.82 / 0.463 & 0.532 / 36.54 / 0.180 \\
Local, random features  & 0.784 / 4.67 / 0.462 & 0.534 / 25.68 / 0.186 \\
Local, one-hot features & 0.779 / 7.25 / 0.452 & 0.518 / 195.59 / 0.182 \\
\bottomrule
\end{tabular}
\end{table}

Table~\ref{tab:feature-ablation} gives two cautions. First, real features can
help: on \textit{dyn\_cora}, the local dataset-feature row has the best local
modularity and NMI, and Table~\ref{tab:stability} shows accelerated
dataset-feature S$^2$CAG on \textit{arxivmath}. Random features are therefore
not a replacement for attributes; they nearly collapse full-graph S$^2$CAG and
become competitive only locally, where topology dominates smaller contracted
instances. Second, one-hot features improve naive modularity on
\textit{dyn\_cora}, but are too expensive on \textit{dyn\_pubmed} and do not
offset their cost locally.
In Table~\ref{tab:closure-ablation}, ``No closure'' keeps only the
radius-expanded vertices, whereas ``No contraction'' gives the backend the
post-closure induced subgraph without community contraction; $T$ is the total
profiled time in seconds.

The direct ablation separates the two roles of the hierarchy. On
\textit{dyn\_cora}, removing closure or contraction can reduce time when
overhead dominates, but no contraction creates an instance about $12\times$
larger than the contracted one and loses modularity. On larger graphs, removing
closure is fastest but loses substantial modularity; removing contraction
keeps quality close on \textit{arxivmath}, but grows the backend instance by
two orders of magnitude. Contraction is therefore what turns closure from
context recovery into a scalable update.

\begin{table}[H]
\centering
\caption{Closure and contraction ablation for Leiden.}
\label{tab:closure-ablation}
\setlength{\tabcolsep}{2pt}
\scriptsize
\begin{tabular}{@{}llrrr@{}}
\toprule
\textbf{Variant} & \textbf{Dataset} & \textbf{$Q$} & \textbf{$T$} &
\textbf{$|\bar V|/|V|$} \\
\midrule
Full ComNetX & dyn\_cora & 0.811 & 0.54 & 0.42\% \\
Full ComNetX & dyn\_pubmed & 0.745 & 0.67 & 0.20\% \\
Full ComNetX & arxivmath & 0.877 & 5.75 & 0.24\% \\
No closure & dyn\_cora & 0.794 & 0.11 & 0.35\% \\
No closure & dyn\_pubmed & 0.565 & 0.76 & 0.18\% \\
No closure & arxivmath & 0.731 & 5.17 & 0.22\% \\
No contraction & dyn\_cora & 0.776 & 0.27 & 4.92\% \\
No contraction & dyn\_pubmed & 0.667 & 10.23 & 8.71\% \\
No contraction & arxivmath & 0.874 & 956.40 & 33.02\% \\
\bottomrule
\end{tabular}
\end{table}

\subsubsection{Real-Data Long-Horizon Robustness}

This experiment addresses the drift component of RQ5.
Protocol P2 evaluates dynamic stability by varying the initial-history fraction
and the number of update batches. Table~\ref{tab:long-horizon} reports paired
$9{:}500$ topology measurements for \textit{dyn\_pubmed} and
\textit{arxivmath}. Rows give final modularity, final-snapshot NMI, cumulative
time, and Local speedup over the corresponding non-local backend.

Local Leiden is about $16\times$ faster than full Leiden on \textit{dyn\_pubmed}
and about $31\times$ faster on \textit{arxivmath}. The NMI gap also grows
relative to the ten-update setting, especially on \textit{dyn\_pubmed}; drift
therefore must be reported explicitly rather than inferred from short streams.

\begin{table}[H]
\centering
\caption{Long-horizon topology endpoints.}
\label{tab:long-horizon}
\setlength{\tabcolsep}{2pt}
\scriptsize
\begin{tabular}{@{}llrrrrr@{}}
\toprule
\textbf{Dataset} & \textbf{Method} & \textbf{Updates} &
\textbf{$Q$} & \textbf{NMI} & \textbf{$T$} & \textbf{Speedup} \\
\midrule
dyn\_pubmed & Leiden & 500 & 0.784 & 0.208 & 315.38 & 1.0$\times$ \\
dyn\_pubmed & Local Leiden & 500 & 0.748 & 0.153 & 19.78 & 15.9$\times$ \\
dyn\_pubmed & DF-Leiden & 500 & 0.762 & 0.195 & 6.23 & 1.0$\times$ \\
dyn\_pubmed & Local DF & 500 & 0.715 & 0.138 & 5.94 & 1.05$\times$ \\
\midrule
arxivmath & Leiden & 500 & 0.887 & 0.427 & 7201.70 & 1.0$\times$ \\
arxivmath & Local Leiden & 500 & 0.874 & 0.355 & 235.84 & 30.5$\times$ \\
arxivmath & DF-Leiden & 500 & 0.855 & 0.388 & 198.86 & 1.0$\times$ \\
arxivmath & Local DF & 500 & 0.869 & 0.335 & 26.84 & 7.4$\times$ \\
\bottomrule
\end{tabular}
\end{table}

\subsubsection{Controlled DSBM Stress Test}

This experiment addresses the failure-mode component of RQ5 by varying update
size and touched-vertex type in an independent DSBM stress suite. Each
synthetic sequence starts from an SBM-style planted-partition snapshot and then
applies controlled edge updates to produce dynamic graph snapshots. Each graph
has $n=100{,}000$ vertices, $k=32$ planted communities, target average degree
$58$, $1\%$ degree-$512$ hubs, and assortativity
$p_{\mathrm{in}}=0.97$. We evaluate random, hub-centered, and
community-internal updates over 100 batches at changed-edge rates of
$0.01\%$ and $0.05\%$, with five seeds for each locality/rate pair.

The DSBM stress study exposes the operating envelope. Table~\ref{tab:dsbm-stress}
summarizes the mean Local speedup and the worst quality loss across five seeds
per setting. At the smaller update rate, all three locality patterns are faster
than full recomputation. At $0.05\%$, random and hub-centered streams fall
below break-even, while community-internal streams remain faster. Larger or
hub-heavy batches should therefore trigger a full refresh.

\begin{table}[H]
\centering
\caption{Controlled DSBM stress summary.}
\label{tab:dsbm-stress}
\setlength{\tabcolsep}{2pt}
\scriptsize
\begin{tabular}{@{}llrrr@{}}
\toprule
\textbf{Rate} & \textbf{Type} & \textbf{Speedup} &
\textbf{Worst $\Delta Q$} & \textbf{Worst $\Delta$NMI} \\
\midrule
0.01\% & random    & 3.34$\times$ & $-$0.0031 & $-$0.0158 \\
0.01\% & hub       & 1.53$\times$ & $-$0.0022 & $-$0.0135 \\
0.01\% & community & 4.29$\times$ & $-$0.0102 & $-$0.0481 \\
0.05\% & random    & 0.89$\times$ & $-$0.0049 & $-$0.0329 \\
0.05\% & hub       & 0.85$\times$ & $-$0.0050 & $-$0.0286 \\
0.05\% & community & 1.13$\times$ & $-$0.0088 & $-$0.0561 \\
\bottomrule
\end{tabular}
\end{table}

\subsubsection{Limitations}

ComNetX targets streams where affected communities remain small relative to the
graph. Hub-heavy or very large updates may approach the full snapshot or lose
boundary context, and representation-learning backends are more sensitive than
Leiden to restricted context and feature representation; monitored workload and
stress-test diagnostics therefore indicate when a full refresh is preferable.
The cross-backend rows use explicit symmetrization for methods that require
undirected graphs; the direction-preserving Leiden experiment supports the same
trend, but directed baselines beyond Leiden and adaptive local resolution remain future
work.

\section{Related work} \label{S:5}
\textbf{Static and modularity-based community detection.} Classical CD methods
rely on modularity, information flow, label propagation, or related structural
criteria. Louvain and Leiden are widely used because they combine high
modularity with strong scalability~\cite{BGLL08,TWE19}, while Infomap and
label-propagation methods offer alternative flow- and propagation-based
views~\cite{RB07,SF09,TS23}. The quality and limitations of such partitions
have been extensively studied, including resolution effects, ground-truth
evaluation, and robustness of detected communities~\cite{POM09,YL15,YAT16,FH16}.
Recent static methods continue to improve modularity optimization through
greedy disassembly and local-move strategies~\cite{Rustamaji24,TTB24}, but
direct application to evolving graphs still tends to require full
recomputation.

\textbf{Dynamic and local community detection.} Dynamic CD methods balance
snapshot quality, temporal consistency, and update efficiency~\cite{RC17,CZW23,HKT17,Sattar23}.
Incremental modularity and label-propagation algorithms update selected parts
of the previous solution~\cite{CT13,SLX14,MZW20,ZCL19,Seif20,XCS13,HLS17}, while
flow- and evolution-based approaches track persistent structures over
time~\cite{Bov22,Sun22,Mazza23}.
Continuous-time objectives, including longitudinal modularity and LAGO, avoid
temporal discretization~\cite{Brabant25}; our setting instead targets batched
snapshot maintenance and reuse of arbitrary solvers under local updates.
Other work studies sparse temporal observations, fully dynamic graph routines,
and scalable multicore or distributed variants~\cite{Conrad25,Blaskovic25,HHC22,SLu24,SKB25,Sattar25}.
Locality appears in several different regimes. Chong and Teow propose batch
incremental maintenance~\cite{CT13}; adaptive label propagation, Delta-Screening,
and dynamic Leiden variants update affected regions through solver-specific
rules~\cite{HLS17,ZK21,SLe24}. DiTursi et al., Christopoulos et al., and Liu
et al. study local, anchor-centered, or seed-based community discovery in
dynamic networks~\cite{DGB17,CBT23,LSS21}, where the output is tied to query or
seed communities rather than a full partition of every snapshot. OLCPM addresses
online overlapping communities~\cite{BCA18}. These methods are important
predecessors for local dynamic processing, but their objectives and output
protocols differ from the solver-agnostic full-partition maintenance studied
here. ComNetX instead wraps arbitrary snapshot and dynamic solvers through
contracted local subproblems and evaluates the same adapter across topology-only
and feature-aware backends.

\textbf{Attributed and GNN-based graph clustering.} Attributed and GNN-based
methods extend CD beyond purely topological criteria. Surveys and recent
models show that neural embeddings can capture high-order dependencies and
combine naturally with modularity or clustering losses~\cite{CCL23,LXW26,TPP23,LYD24,ZXZ20}.
Examples include modularity-oriented contrastive learning and neural modularity
maximization~\cite{LLC24,BKH24}, structural-entropy objectives~\cite{ZPS25},
pre-train-and-refine pipelines for scalable partitioning~\cite{QZG24,QZG24HPEC},
and spectral subspace clustering for attributed graphs~\cite{LYZ25}. Recent
models also explore fairness-aware deep community detection~\cite{GPT25}.
Although these methods improve expressiveness, they are often expensive to
rerun after each graph update.

Overall, prior work offers either efficient routines specialized to a fixed
dynamic design or expressive static and feature-aware solvers that are costly to
rerun. ComNetX is complementary: it wraps static, GNN-based, and native dynamic
solvers while restricting computation to affected aggregated subproblems.
\section{Conclusion} \label{S:6}
We studied the trade-off between reusable full-snapshot community detection and
efficient solver-specific dynamic updates. Full recomputation preserves solver
semantics but revisits unchanged graph regions; specialized dynamic algorithms
reduce update cost but are less transferable across objectives and
implementations; and purely neighborhood-based localization can lose the
community context required by strong solvers. ComNetX addresses these
limitations through a solver-agnostic hierarchical adaptation mechanism: it
preserves a community hierarchy, closes each affected region over the
communities that provide context, contracts that region into a compact backend
instance, and projects the updated labels back to the full graph.
The main contribution is a reusable local-update layer that allows
topology-only, feature-aware, and native dynamic solvers to operate on
contracted affected subproblems through a common interface. The experiments show
that this design preserves much of the modularity of full recomputation for
Local Leiden while reducing update time on larger networks; ablations show that
radius, hierarchy depth, and feature representation materially affect the
trade-off; and long-horizon and DSBM stress tests show when cumulative speedups
persist or a full refresh is preferable.
\begingroup
\setlength{\labelsep}{0.12em}

\endgroup

\end{document}